\documentstyle[12pt]{article}
\newcommand{\nc}{\newcommand}
\nc{\be}{\begin{equation}}
\nc{\ee}{\end{equation}}
\nc{\ba}{\begin{eqnarray}}
\nc{\ea}{\end{eqnarray}}
\begin{document}
\begin{titlepage}
\rightline{TURKU - FL - R18}
\vskip 3cm

\begin{center}
\bigskip

{\LARGE \bf Baryogenesis in the singlet Majoron model}
\bigskip
\bigskip

I. Vilja\footnote{internet: vilja@sara.cc.utu.fi}\\
Department of Physics\\
University of Turku\\
FIN-20500 Turku\\
Finland\\
%\bigskip
%and
%somebody
\bigskip\bigskip\bigskip
{\bf Abstract}
\end{center}
\begin{quote}
\baselineskip 0.7cm

The baryogenesis by decays of the right--handed neutrinos in the Majoron model
is
studied. We show that compared to the observational value
it is possible to produce large enough baryon asymmetry in the Majoron
model or, at least, contribute strongly to it.
It requires that the mass of the lightest right--handed neutrino is
about (or above) 1 TeV and the self coupling of the singlet sector is weak
enough,
$ < 2\times 10^{-5}$, to prevent the thermalization of the right--handed
neutrinos
by annihilations.
\end{quote}
\vfill
\end{titlepage}

\baselineskip 0.7cm

The singlet Majoron model \cite{cmp} is one of the simplest extensions of the
Standard
Model explaining many question remained open in the Standanrd Model.
In the context of the Majoron model the vanishingly small neutrino masses can
be
explained by using spontaneous breaking of a global U(1) symmetry and sew--saw
mechanism. The right handed neutrinos aquire large Majorana masses while the
left handed ones keep very light. On the other hand, in the Standard Model the
baryogenesis appears to be difficult to realize \cite{krs,Sha} due to too small
CP--violation and sphaleron \cite{km} mediated $B+L$ violating transitions.
Those transitions not only
erase any pre--existing baryon asymmetry, but are also able to wash out the
asymmetry produced  during the electro--weak phase transition. In the matter of
fact, the experimental Standard Model higgs mass lower bound $m_H> 60$ GeV
\cite{Mori}
indicates that the observed
baryon--to-photon ratio $n_B/n_\gamma \sim 10^{-10}$ is too large to be
explained by baryogenesis during the electroweak phase transition of the
minimal Standard
Model.

It has been proposed that the decay of heavy Majorana neutrinos could be
responsible of the baryogenesis \cite{fy&ckn,Vay,ev2}. In this scenario the
out--off--equilibrium
decays of heavy right--handed neutrinos produce a non--zero lepton
number $L$ which is converted to  $B$ asymmetry by sphaleron mediated
transitions. Therefore the singlet Majoron model, as a minimal extension of the
Standard Model, is a natural and interesting candidate for
explaining both the neutrino masses and the observed baryon asymmetry. The
relation
between baryogenesis scale and the neutrino masses has also been discussed in
the
literature \cite{several}. In this paper we
study the singlet Majoron model, in particular the decay rate of the heavy
right--handed neutrinos and compare it to their
annihilation rate showing that for suitable parameters the baryon
asymmetry can be produced although it seems to require that the singlet
sector is weakly coupled to the the doublet sector. For
these purposes we have to study the phase structure of the Majoron model and
equilibration of the right--handed neutrinos.

The Majoron model contains, in addition to the Standard Model doublet Higgs
$H$,
one electroweak singlet field $S$. To the fermionic sector has been added the
right--handed neutrinos $N_i$, where $i$ is a generation index. Both $S$ and
$N_i$
are in non--trivial representations of the global U(1) group, whereas other
fermions and the Higgs field $H$ are singlets with respect it. The classical
potential reads
\be
V_0(H,S) = m_H^2|H|^2 + m_S |S|^2 + \gamma |H|^2|S|^2 + \beta |S|^4
+ \lambda |H|^4 \label{pottree}
\ee
where the mass--like parameters $m_H^2,\ m_S^2 < 0$ and the couplings $
 \beta,\ \lambda > 0$ with $\gamma^2 <
4\lambda\beta$. (See Ref. \cite{ekv} for detailed study of the potential.)
At finite temperature the potential is modified at one--loop level (keeping
only leading terms) as
\be
V_T(H,S) = \mu_H^2(T)|H|^2 + \mu_S^2(T)|S|^2 + \gamma |S|^2|H|^2 +
\beta |S|^4 + \lambda |H|^4 - {T\over 12\pi}\mbox{\rm tr} M^3 +...,\label{potT}
\ee
where $M$ is the general scalar mass matrix and the dots stand for remaining
finite temperature corrections. The temperature corrected masses introduced in
(\ref{potT}) are given by
\ba
\mu_H^2(T)&= &m_H^2 + r'\frac {T^2}4,\\
\mu_S^2(T)&=& m_S^2 + s'\frac {T^2}3,\label{mus}
\ea
where
\ba
r'&=& {2m_W^2+m_Z^2+2m_t^2\over f^2} \simeq 0.67,\label{r'}\\
s'&=& \beta + \frac 18 \gamma + \frac 18 h_\nu^2,\label{s'}
\ea
$f=247$ GeV is the vacuum expectation value of the doublet higgs and $h_\nu^2$
stands collectively for all right--handed Yukawa couplings: $h_\nu^2 = \sum_i
h_i^2$. In the equation (\ref{r'}) we have omitted the scalar
contribution to $r'$ as small. The cubic term $\mbox{\rm tr} M^3 $ is a
complicated
expression of the fields \cite{ekv}, but can be simplified considerably in
some special cases.

To make the baryogenesis possible, the symmetry breaking has to proceed first
to
the direction of the singlet field $S$, so that only the global U(1) symmetry
breaks generating  large Majorana masses for the right--handed neutrinos.
 The ratio of critical temperatures
in the directions of $S$ and $H$, $T_{c,S}$ and $T_{c, H}$, respectively, are
given by
\be
\left ( {T_{c,S}\over T_{c,H}} \right )^2 = {m_\phi^2\over m_h^2}{\frac 14 r' -
\frac 2{9 \lambda} \left ({2r\over 4\pi} + {\gamma^{3/2}\over 4\pi\sqrt 2}
\right )^2\over \frac \beta 3 + \frac {\gamma + h_\nu^2} 8 \left (\frac 13 -
 \frac{\sqrt{r'}}{2\pi} \right ) - \frac 29 ({1 + 3^{3/2}\over
4\pi})^2\beta^2},\label{ratio}
\ee
where $r={2m_W^3+m_Z^3\over f^3}\simeq 0.35$. The phase structure of the
singlet
Majoron model  has been extensively analyzed in Ref. \cite{ekv}. The ratio
(\ref{ratio}) has to be larger than one
in order to have two--stage phase transition proceeding first to the singlet
direction. If U(1)--symmetry is broken first, the critical temperature to the
doublet direction is
\be
(T'_{c,H})^2 = 4\, {m_H^2 - \frac \gamma{2\beta} m_S^2\over r' -
\frac{r^2}{2\pi^2\lambda}}.
\ee

It is of essential importance that the right--handed  neutrinos have also
obtained
masses before the sphaleron transitions freeze out at $T_{sph}$, i.e. before
 they become too
ineffective to convert any lepton number to baryon number. We require that the
mass $m_{N_1} = h_1 \bar f$ of the lightest right--handed neutrino is
larger than $T_{sph}$ given by solving the equation \cite{Sha}
\be
E_{sph}^{SM}(T_{sph})/T_{sph} \equiv x \simeq 45,\label{tspha}
\ee
where $E_{sph}^{SM}(T) = 4\pi B(\lambda)f(T)/g_W$ is the Standard Model
sphaleron energy, $g_W^2 = 0.436$ and
 $B(\lambda)$ is a smooth, increasing function varying between
values 1.56 and 2.72. Although Eq. (\ref{tspha}) defines the
freeze--out temperature for the Standard Model, it is with high accurary same
 than in the Majoron model \cite{ev}. Thus ${m_{N_1}\over T_{sph}} > 1$ is a
 necessary condition for baryogenesis via neutrino decay. It can be cast in the
form
\be
1 < {m_{N_1}^2\over T_{sph}^2} = \left ({2 r\over4\pi}\right )^2\left [ \left (
{2 x g_W\over 3 r B(\lambda)}\right )^2\lambda + r'\left ( {4\pi\over 3r}\right
)^2 - {4 x g_W\over 3 r B(\lambda)}\right ] {4\beta\over 4\beta\lambda -
\gamma^2} {m_{N_1}^2\over f^2}.\label{c1}
\ee

 The decay rate of the lightest right--handed neutrino (chosen to be the first
generation neutrino) is given by
\be
\Gamma_d = {1\over 8\pi^2}{m_{d_1}^2\over f^2}m_{N_1},\label{drate}
\ee
where $m_{d_1}$ is the Dirac mass of the lightest generation, $m_{d_1} \simeq
m_e \sim 1$ MeV. This rate has to be compared to the expansion rate of the
universe,
to the Hubble  rate $H$:
\be
H(T) = 1.66 g_*^{1/2}{T^2\over M_{Pl}},\label{hrate}
\ee
where $g_*$ is the effective number of degrees of freedom, $g_* = 110.5$
 (at $T\sim m_{N_1}$), and
$M_{Pl} \simeq 1.2\, 10^{19}$ GeV is the Planck mass. A condition for
out--of--equilibrium decay can be written now as
\be
\Gamma_d < 2 H\label{c2}
\ee
ehich has to be evaluated at $T\sim m_{N_1}$. This is, however, not a
sufficient condition,
 but one has to take
accout that the annihilations $NN \rightarrow \chi\rho$ and $ NN\rightarrow
\chi\chi$,
 where $\chi$
is the Majoron and $\rho$ is the massive part of the singlet scalar, tend to
keep the right--handed neutrinos in thermal equilibrium. The thermally averaged
annihilation rates are approximately given by
\be
\Gamma_{ann} = \langle \sigma v_{rel}\rangle n(T) \simeq {\beta^2\over 32\pi}
\left ( 1-
{m_\rho^2\over 4 m_N^2}\right )^{\pm 1}{n(T)\over m_N^2},\label{arate}
\ee
where $m_\rho^2 = 2\beta \bar f^2$ is the singlet scalar mass  before
electroweak
breaking and  $n(T)$ is the usual number density of the right--handed
neutrinos. The upper sign refers to the process $NN \rightarrow \chi\rho$
whereas
 the lower sign refers to the process $ NN\rightarrow \chi\chi$.
At the relevant temperature, $T \sim m_{N_1}$, the annihilation rate
is $ \Gamma_{ann} \simeq 4.6\, 10^{-4}\beta^2m_{N_1}(1 -
\frac{m_\rho^2}{4m_{N_1}})^{\pm 1}$. We require that the annihilations must
proceed
much slower than the decays, i.e. $\Gamma_{ann} \ll \Gamma_d$.  Explicitely
written
the condition reads
\be
\beta\left (1 - {\beta\over 2 h_1^2}\right )^{\pm 1/2} \ll 2\times
10^{-5}.\label{c3}
\ee

We have got three necessary conditions (\ref{c1}), (\ref{c2}) and (\ref{c3})
for
 non-equilibrium decays of the right--handed neutrinos.
 It as fairly easy to see that the
condition (\ref{c1}) is practically allways satisfied if  the condition
(\ref{c2}) is, and the condition (\ref{c3}) implies $\beta \ll 2\times
10^{-5}$.
On the other hand in order to have that small
value of $\beta$ stable against radiative effects, in particular the correction
from $H$--loop
has to be small enough, i.e. ${\gamma^2\over 64\pi^2} \ll 2\times 10^{-5}$.
This yields
an upper bound  $\gamma \ll 0.12$. In practise, however, this bound is not very
restictive,
but the stability bound of the potential, $\gamma^2 < 4\lambda\beta$, is much
stricter.

 The analysis above shows, when the enviroment for the baryon asymmetry
generation is
favorable, but it does not tell anything precise about
 the net baryon number $N_B = n_B/n_\gamma$ produced in the neutrino
decays. For that purpose one has to calculate first the $B - L$ number,
$N_{B-L}$, produced
by decays of the right--handed neutrinos. This can be done  by
solving the relevant Boltzman equation,  yielding  \cite{kt}
\be
N_{B-L} \simeq {0.3 \epsilon\over g_* K(\ln K)^{0.6}},\label{nb}
\ee
where
\ba
K &=& {\Gamma_d(T)\over 2 H(T)}\Bigg |_{T = m_{N_1}},\\
\epsilon &\simeq& \frac 1\pi {m_{D_3}^2\over f^2}{m_{N_1}\over m_{N_3}}\delta
\ea
and $\delta$ is the CP--violating phase. Here is assumed that $h_3 \gg h_2\gg
h_1$,
i.e. the third generation is the heaviest and the first generation is the
lightest one.
Afterwards the lepton number $N_L$ ($= N_{B-L}$ here) is convereted to the
baryon number
 $N_B$ by
the sphaleron mediated transitions according to \cite{Vay}
\be
N_B = {28\over79} N_{B-L}.
\ee

To get some idea about the magnitude of the baryon number generated, we use
values
like $m_{D_1}\sim 1$ MeV, $m_{D_3}\sim 1$ GeV, $m_{N_1} \sim 1$ TeV and
${m_{N_1}\over m_{N_3}}\delta\sim 10^{-1}$. We obtain $N_B\sim 10^{-10}\ -\
10^{-12}$ and
clearly the right--handed neutrino decays are at least contributing to the
baryon number
of the universe. It is, however, clear that the calculation contains many
uncertainties,
 primarily due to uncertainties on neutrino masses. In particular, if $m_{N_1}
< 1$ TeV,
the right--handed neutrino contribution would be smaller than $10^{-12}$. On
the other hand,
if $m_{N_1}$ were larger than few TeV's, the appealing feature of the dynamical
contact
 between U(1) breaking and electroweak breaking would be lost, unless the
paremeters of the
potential are suitably fine--tuned \cite{ekv}.
The result of the present paper, together with other
baryogenesis results of the singlet Majoron model \cite{ekv}, suggests anyway
that the
observed baryon asymmetry is, perhaps, not a result of a single baryogenesis
mechanism
but might be produced as result of several physical phenomena.

\end{document}